\documentstyle[twocolumn,epsf]{jpsj}

\def\runtitle{Vortex Structure in Superconducting Stripe States }

\def\runauthor{
Masanori {\sc Ichioka},  Mitsuaki {\sc Takigawa} and Kazushige {\sc Machida}
}

\title{Vortex Structure in Superconducting Stripe States }

\author{
Masanori {\sc Ichioka}\footnote{E-mail address: oka@mp.okayama-u.ac.jp},
Mitsuaki {\sc Takigawa}  and Kazushige {\sc Machida}
}

\inst{Department of Physics, Okayama University, Okayama 700-8530}

\recdate{August 14, 2000}

\abst{
The vortex structure in superconducting stripe states is studied 
according to the Bogoliubov-de Gennes theory on the two-dimensional 
Hubbard model with nearest-neighbor sites pairing interaction. 
The vortex is trapped at the outside region of the stripe line, 
where the superconductivity is weak. 
The superconducting coherence length along the stripe direction 
becomes long. 
There are no eminent low-energy electronic states even near the vortex core. 
These characters resemble the Josephson vortex in layered superconductors 
under a parallel field. 
}

\kword{
vortex structure, 
stripe state, 
local density of states, 
Bogoliubov-de Gennes theory 
}

\begin{document}

%

\sloppy
\maketitle


\fulltext
\halftext


Recently, much attention has been focused on the stripe state of  underdoped
high-$T_{\rm c}$ cuprates.
The stripe state was proposed to explain the static magnetic incommensurate 
structure observed by means of elastic neutron scattering experiments on
La$_{2-x}$Sr$_x$CuO$_4$ (LSCO)\cite{Matsuda,Wakimoto,Suzuki} and 
La$_{1.6-x}$Nd$_{0.4}$Sr$_x$CuO$_4$ (LNSCO).\cite{TranquadaR,TranquadaNd} 
It is considered that doped holes are localized in the stripe region,
which contributes to the one-dimensional (1D) metallic conduction,\cite{Noda}
and the outside region of the stripe is an antiferromagnetic (AF) insulator.
Angle-resolved photoemission (ARPES) experiments were carried and 
the 1D-like Fermi surface in the stripe state was observed.\cite{Zhou,Ino1,Ino2,Ino3}  
In ${\rm Y Ba_2 Cu_3 O_{7-\delta}}$, incommensurate fluctuations that were 
consistent with the above stripe concept were reported by means of 
inelastic neutron scattering experiments.\cite{Mook1,Mook2,Arai}

In high-$T_{\rm c}$ superconductors, it is considered that the low-energy 
electronic state around the vortex is completely different from that of 
conventional superconductors.
Theoretical studies\cite{Volovik,IchiokaDL1,IchiokaDL2,Wang,Franz} 
suggest that the low-energy electronic state around the 
vortex core extends a significant distance due to the line node of 
the $d$-wave superconducting gap in high-$T_{\rm c}$ superconductors. 
Thus, we expect the zero-energy peak in the local density of states (LDOS) 
at the vortex core, instead of the quantized energy level of the 
conventional $s$-wave pairing case. 
However, in the direct observation of the vortex core by scanning 
tunneling microscopy (STM), there is no eminent low-energy 
state around the vortex core. 
There appears only a small unexplained shoulder or isolated peak 
at a higher energy within the superconducting gap.\cite{pan,Renner,Maggio1,Maggio2}  
These results suggest that we must consider the vortex state, 
including the exotic character of the electronic state which is unique  
to high-$T_{\rm c}$ materials.\cite{Himeda,Ogata,Han} 
We study the effect of the stripe state in this paper. 

Although there are many theoretical approaches to the stripe state, 
we base this study on the self-consistent Hartree-Fock (HF) theory of 
the Hubbard model. 
It is believed that the stripe concept is valid beyond the HF 
approximation.\cite{Zaanen} 
We can consider the metallic stripe state
by using the self-consistent HF theory if we consider the
realistic Fermi surface topology.\cite{MachidaL,Ichioka}
It can reproduce ``the 1D Fermi surface with the gap near
$(\frac{\pi}{2},\frac{\pi}{2})$''
[see Fig. 12(d) in ref. \citen{Ichioka}],
which is suggested by the results of ARPES 
experiments.\cite{Zhou,Ino1,Ino2,Ino3}
It also qualitatively reproduces the relationship between the 
incommensurability and the hole density $n_{\rm h}$, including the phase 
transition between the diagonal stripe in the insulator phase 
($n_{\rm h}<0.05$) and the vertical stripe in the metallic and superconducting 
phases ($n_{\rm h}>0.05$).\cite{Matsuda,Wakimoto,Suzuki}
Thus, HF theory can successfully describe 
the stripe structure in high-$T_{\rm c}$ cuprates as a first approximation.
Therefore, we further investigate various phenomena in the superconducting 
state by extending this theory.  
Once the density of states (DOS) remains at the Fermi energy 
(i.e., metallic state), we can produce the superconductivity 
by introducing the pairing interaction, at least as a phenomenological 
model.\cite{IchiokaC,Martin} 
In the superconducting state of this framework, the metallic 1D stripe region 
becomes dominantly superconducting, 
and penetrates the outside AF region.  
In this paper, we investigate the vortex state under a magnetic field 
in this superconducting state, and study the structure of the order parameter 
and the electronic state around the vortex. 

We begin with the conventional Hubbard model on a two-dimensional
square lattice, and introduce the mean field
$ n_{i,\sigma}= \langle a^\dagger_{i,\sigma} a_{i,\sigma} \rangle$
at the $i$-site, where $\sigma$ is a spin index and $i=(i_x,i_y)$. 
We assume a pairing interaction $V$ between nearest-neighbor (NN) sites. 
This type of pairing interaction gives the $d$-wave 
superconductivity.\cite{Wang}  
Thus, the HF Hamiltonian under a magnetic field is given by 
\begin{eqnarray}
{\cal H}&=&
-\sum_{i,j,\sigma} \tilde{t}_{i,j}a^{\dagger}_{i,\sigma} a_{j,\sigma}
+U\sum_{i,\sigma} n_{i,-\sigma}
a^\dagger_{i,\sigma} a_{i,\sigma} 
\nonumber \\ &&
+V\sum_{\hat{e},i,\sigma}
(\Delta^\dagger_{\hat{e},i,\sigma} a_{i,-\sigma} a_{i+\hat{e},\sigma}
+\Delta_{\hat{e},i,\sigma} a^\dagger_{i,\sigma} a^\dagger_{i+\hat{e},-\sigma}
) , \qquad 
\label{eq:2.2}
\end{eqnarray}
where $a^{\dagger}_{i,\sigma}$
($a_{i,\sigma}$) is a creation (annihilation) operator, and 
$i+\hat{e}$ represents the NN site ($\hat{e}=\pm\hat{x},\pm\hat{y}$).
The transfer integral is expressed as 
\begin{equation}
\tilde{t}_{i,j}=t_{i,j} \exp [ {\rm i}\frac{\pi}{\phi_0}\int_{{\mib
r}_i}^{{\mib r}_j} {\mib A}({\mib r}) \cdot {\rm d}{\mib r} ] , 
\label{eq:BdG4}
\end{equation}
with the vector potential 
${\mib A}({\mib r})=\frac{1}{2}{\mib H}\times{\mib r}$ in the symmetric gauge, 
and the flux quantum $\phi_0$.
For the NN pairs $(i,j)$, $t_{i,j}=t$.
For the next-NN pairs situated on a diagonal position on the square
lattice, $t_{i,j}=t'$. 
For the third-NN pairs, which are situated along the NN bond
direction, $t_{i,j}=t''$.
To reproduce the Fermi surface topology of LSCO, we set $t'=-0.12t$ and 
$t''=0.08t$.\cite{Tohyama} 
The essential results of this paper do not significantly depend on the 
choice of these parameter values. 

In terms of the eigen-energy $E_\alpha$ and the wave functions
$u_\alpha({\mib r}_i)$, $v_\alpha({\mib r}_i)$ at the $i$-site,
the Bogoliubov-de Gennes equation is given by
\begin{equation}
\sum_j
\left( \begin{array}{cc}
K_{\uparrow,i,j} & D_{i,j} \\ D^\dagger_{i,j} & -K^\ast_{\downarrow,i,j}
\end{array} \right)
\left( \begin{array}{c} u_\alpha({\mib r}_j) \\ v_\alpha({\mib r}_j)
\end{array}\right)
=E_\alpha
\left( \begin{array}{c} u_\alpha({\mib r}_i) \\ v_\alpha({\mib r}_i)
\end{array}\right) ,
\label{eq:BdG1}
\end{equation}
where
$K_{\sigma,i,j}=-\tilde{t}_{i,j} +\delta_{i,j} (U n_{i,-\sigma} -\mu)$, 
$D_{i,j}=V \sum_{\hat{e}}  \Delta_{i,j} \delta_{j,i+\hat{e}} $ 
and $\alpha$ is an index of the eigenstate.\cite{Wang,Takigawa}   
The self-consistent condition for the pair potential 
and the number density is given by 
\begin{eqnarray} &&
\Delta_{i,j}
=\langle a_{j,\downarrow} a_{i,\uparrow} \rangle 
=\sum_\alpha  u_\alpha({\mib r}_i) v^\ast_\alpha({\mib r}_j) 
f(E_\alpha) , 
\label{eq:BdGsc1} \\&&
n_{i,\uparrow}=\langle a^\dagger_{i,\uparrow} a_{i,\uparrow} \rangle 
=\sum_\alpha |u_\alpha({\mib r}_i)|^2 f(E_\alpha ), 
\label{eq:BdGn1} \\&&
n_{i,\downarrow}=\langle a^\dagger_{i,\downarrow} a_{i,\downarrow}  \rangle 
=\sum_\alpha |v_\alpha({\mib r}_i)|^2 (1-f(E_\alpha )) . 
\label{eq:BdGn2}
\end{eqnarray}
The charge density $n_i=n_{i,\uparrow}+n_{i,\downarrow}$ and 
the spin density $S_{z,i}=\frac{1}{2}(n_{i,\uparrow}-n_{i,\downarrow})$. 
The superconducting order parameter is decomposed into  
$d$- and  $s$-wave components at each site $i$ as
\begin{eqnarray} &&
\Delta_{d,\sigma,i}=
( \Delta_{ \hat{x},i,\sigma}
 +\Delta_{-\hat{x},i,\sigma}
 -\Delta_{ \hat{y},i,\sigma}
 -\Delta_{-\hat{y},i,\sigma} )/4,
\\ &&
\Delta_{s,\sigma,i}=
( \Delta_{ \hat{x},i,\sigma}
 +\Delta_{-\hat{x},i,\sigma}
 +\Delta_{ \hat{y},i,\sigma}
 +\Delta_{-\hat{y},i,\sigma} )/4  , 
\end{eqnarray}
with
\begin{equation}
\Delta_{\hat{e},i,\sigma}=\Delta_{i,i + \hat{e},\sigma}
\exp[{\rm i}\frac{\pi}{\phi_0}
\int_{{\mib r}_i}^{({\mib r}_i+{\mib r}_{i + \hat{e}})/2}
{\mib A}({\mib r}) \cdot {\rm d}{\mib r}], 
\label{eq:dOP2}
\end{equation}
where 
$\Delta_{i,i+\hat{e},\uparrow}=\langle a_{i+\hat{e},\downarrow} 
a_{i,\uparrow} \rangle$ and 
$\Delta_{i,i+\hat{e},\downarrow}=- \langle a_{i+\hat{e},\uparrow} 
a_{i,\downarrow} \rangle$.

We typically consider the case of a unit cell with $24 \times 24$ sites, 
where two vortices are accommodated.
Regarding the stripe structure, 
we assume a vertical stripe of the incommensurability $\delta=\frac{1}{8}$, 
i.e., an eight-site periodic spin structure. 
It is characterized by the ordering vector 
${\mib Q}=2\pi (\frac{1}{2},\frac{1}{2}-\delta)$, 
where the lattice constant is unity.  
The spatially averaged hole density is set to 
$n_{\rm h}=1-\overline{n_i} \sim \frac{1}{8}$ by tuning the 
chemical potential $\mu$. 
By introducing the quasimomentum of the magnetic Bloch state,
we obtain the wave function under the periodic boundary condition 
whose region covers many unit cells. 
We consider the low-temperature case $T=0.01t$, where the vortex structure 
is almost independent of $T$. 

\begin{fullfigure}
\begin{center} 
\leavevmode
\epsfbox{IKMfig1.epsi}
\end{center}
\caption{
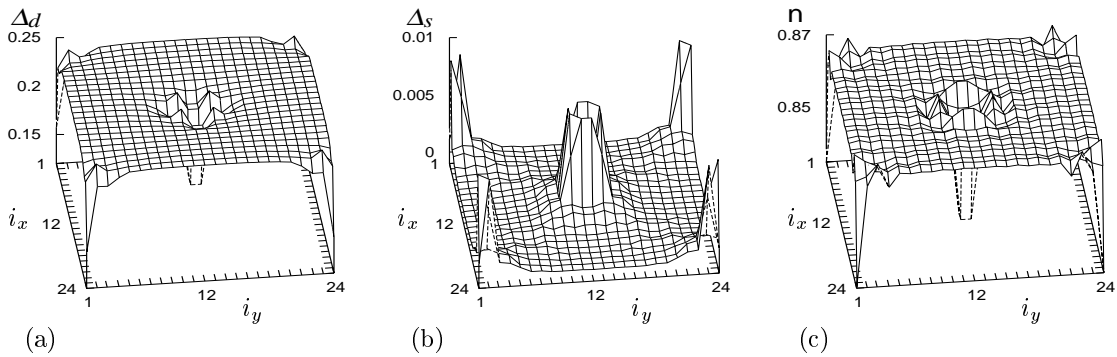
Vortex structure in the no-stripe case for $U=0$ and $V=-2t$. 
(a) $|\Delta_{d,i}|$, (b) $|\Delta_{s,i}|$, (c) $n_i$. 
The area of $24 \times 24$ sites is plotted, where 
vortices are located in the middle and at the four corners of the figure. 
}
\label{fig:1}
\end{fullfigure}

First, we discuss the spatial structure of the order parameter 
for the $d$-wave ($s$-wave) superconductivity
$\Delta_{d,i}=\Delta_{d,i,\uparrow}+\Delta_{d,i,\downarrow}$ 
($\Delta_{s,i}=\Delta_{s,i,\uparrow}+\Delta_{s,i,\downarrow}$) 
and the charge density $n_i$ by comparing the usual no-stripe case  
and the stripe case. 
Figure \ref{fig:1} presents the vortex structure in the no-stripe case 
$U=0$ and $V=-2t$. 
The amplitude $|\Delta_{d,i}|$ in Fig.  \ref{fig:1}(a) is suppressed near 
vortices. 
The vortex centers are located in the middle and at the four corners of the 
figure.  
Around the vortex core, the $s$-wave component $\Delta_{s,i}$ is induced,  
as shown in Fig.  \ref{fig:1}(b), 
because the fourfold symmetry of the system is locally broken 
where the dominant $d$-wave order parameter varies spatially. 
Figure \ref{fig:1}(c) shows that the charge density $n_i$ is slightly 
suppressed in the vortex core region. 
This is a charging effect of the vortex core, which has been discussed 
in the $s$-wave pairing case.\cite{Khomskii,Hayashi}  
We show that this occurs even in the $d$-wave pairing case by means of the 
microscopic calculation. 
It is suggested that the charging effect is related to 
the Hall anomaly in high-$T_{\rm c}$ superconductors.\cite{Nagaoka}   
However, to explain the experimental results, we need the opposite sign of the 
charging to those theoretically obtained thus far. 
We obtain the opposite sign in the stripe state, as mentioned later. 

The stripe structure without a vortex was reported in refs. 
\citen{Ichioka}, \citen{IchiokaC}  and \citen{Martin}. 
For finite dopings from half-filling, the AF structure is modulated. 
For $\delta=\frac{1}{8}$, the envelope of the AF moment 
$(-1)^{i_x+i_y}S_{z,i}$ shows the wave of the eight-site 
period along the $y$- (or $x$-) direction. 
Near the stripe line where  $(-1)^{i_x+i_y}S_{z,i}$  
changes its sign and the $\pi$-shift occurs, 
the magnetic moment is suppressed. 
Along the stripe line, the doped carriers (holes) accumulate, 
and the $d$-wave superconductivity $|\Delta_{d,i}|$ is large. 
Outside of the stripe, $|\Delta_{d,i}|$ is suppressed. 

\begin{fullfigure}
\begin{center}
\leavevmode
\epsfbox{IKMfig2.epsi}
\end{center}
\caption{
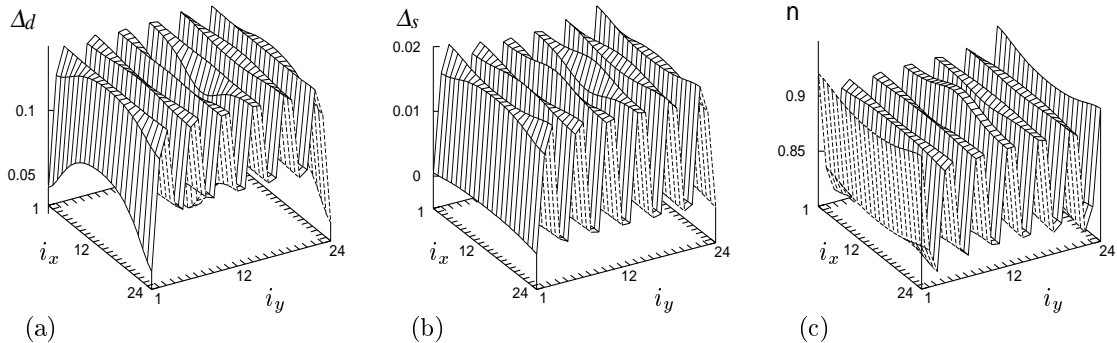
Vortex structure in the stripe state for $U=4t$ and $V=-2t$.  
(a) $|\Delta_{d,i}|$, (b) $|\Delta_{s,i}|$, (c) $n_i$. 
The area of $24 \times 24$ sites is plotted, where 
vortices are located in the middle and at the four corners of the figure. 
Bond-centered stripe lines run along the $x$ direction, 
which are at the bond $i_y$=2-3, 6-7, 10-11, 14-15, 18-19, 22-23. 
}
\label{fig:2}
\end{fullfigure}

Figure \ref{fig:2} shows the vortex structure in the stripe state for 
$U=4t$ and $V=-2t$. 
Figure \ref{fig:2}(a) is for $|\Delta_{d,i}|$. 
Here, we show the bond-centered stripe case, where 
the center of the stripe line is on a bond. 
The site with large $|\Delta_{d,i}|$ is in the stripe region, and 
that with small $|\Delta_{d,i}|$ is outside of the stripe. 
While  $|\Delta_{d,i}|\sim 0.24$ in the no-stripe case 
in Fig. \ref{fig:1}(a), $|\Delta_{d,i}|$ is suppressed to 
$|\Delta_{d,i}|\sim 0.15$ at its maximum. 
This is because part of the DOS at the Fermi energy is 
already used for the stripe formation, and the superconductivity occurs 
by using the remaining DOS. 
The vortex center tends to be attracted to the outside of the stripe, 
where $|\Delta_{d,i}|$ is weak. 
We consider various cases for the vortex center position as an initial 
state of the iterating calculation. 
However, the vortex center is shifted to the outside of the stripe in the 
final self-consistent results.  
This suggests that the stripe functions as a line pinning center for 
a vortex.  
This is reasonable because the condensation energy loss required to create 
a vortex is minimal. 
Figure \ref{fig:2}(a) shows that $|\Delta_{d,i}|$ suppressed 
near the vortex core recovers quickly in the direction perpendicular 
to the stripe. 
The effect of the vortex is limited to within about three sites 
from the vortex center. 
However, along the parallel direction, $|\Delta_{d,i}|$ around the vortex 
core slowly recovers by using the maximal length of the intervortex distance. 
That is, the superconducting coherence length is short (long) 
in the perpendicular (parallel) direction. 
This suggests that the stable vortex lattice configuration is modified 
from the conventional $60^\circ$ triangular lattice. 
However, the vortex core structure, which we are studying, will not be 
significantly affected by the vortex lattice configuration.

Figure \ref{fig:2}(b) shows the small $s$-wave component  $|\Delta_{s,i}|$ 
induced by the spatial variation of the dominant $|\Delta_{d,i}|$. 
There, the effect of the stripe is large compared with the vortex effect. 
In the stripe region, $|\Delta_{s,i}|$ is large, 
and $|\Delta_{s,i}|$ is slightly suppressed near the vortex. 
As shown in Fig. \ref{fig:2}(c), $n_i$ is large  outside of the stripe. 
There, $n_i$ is enhanced at the vortex core. 
It is important to note that the sign of the charging is opposite 
to that of the no-stripe case. 
This sign change is related to the AF moment of the stripe state. 
Also, when the AF moment is induced around the vortex, this type of 
sign change is reported.\cite{Ogata}  
The structure of $S_{z,i}$ is not strongly affected by the vortex 
in our results.  

Next, we investigate the electronic structure around the vortex 
by calculating the LDOS 
$N(E,{\mib r}_i)=N_{\uparrow}(E,{\mib r}_i)+N_{\downarrow}(E,{\mib r}_i)$ 
at the $i$-site, where $N_{\uparrow}(E,{\mib r}_i)
=\sum_\alpha  |u_\alpha({\mib{r}})|^2\delta(E-E_\alpha)$ for up-spin and 
$N_{\downarrow}(E,{\mib r}_i)
=\sum_\alpha  |v_\alpha({\mib{r}})|^2\delta(E+E_\alpha)$ for
down-spin contributions. 
At the site where $S_{z,i}>0$, 
$N_{\uparrow}(E,{\mib r}_i)>N_{\downarrow}(E,{\mib r}_i)$ for $E<E_{\rm F}$. 
In our numerical calculation, we use 
the derivative $f'(E)$ of the Fermi function instead of $\delta(E)$. 
In this case, $N(E,{\mib{r}})$  corresponds to the differential
tunnel conductance of STM experiments.

\begin{fullfigure}
\begin{center}
\leavevmode
\epsfbox{IKMfig3.epsi}
\end{center}
\caption{
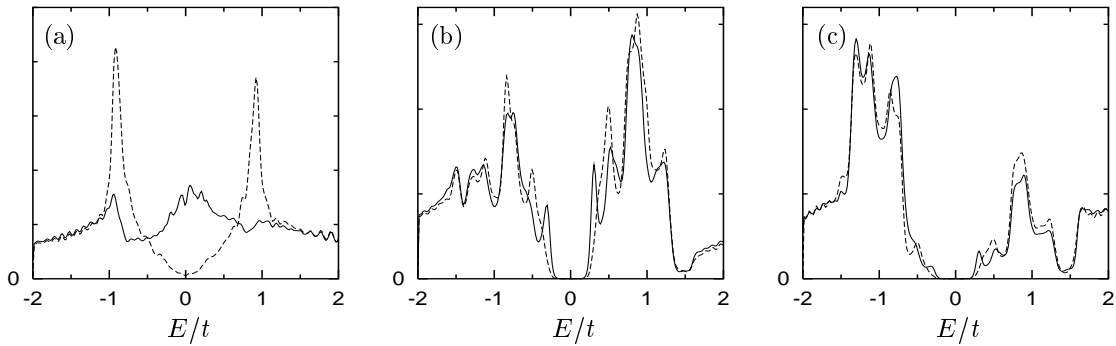
Local density of states $N(E,{\mib r}_i)$ at the vortex core (solid lines) 
and far from the vortex (dotted lines). 
(a) 
The no-stripe case of Fig. \protect{\ref{fig:1}}. 
We plot  $N(E,{\mib r}_i)$ at the site $(i_x,i_y)=(1,1)$ for the vortex core, 
and at the $(12,1)$-site far from the vortex.  
(b) 
The stripe region in the stripe case of Fig. \protect{\ref{fig:2}}. 
We plot  $N(E,{\mib r}_i)$ at the  $(1,2)$-site in the vortex core, 
and at the $(12,2)$-site far from the vortex.  
(c) 
The outside region of the stripe in the stripe case of 
Fig. \protect{\ref{fig:2}}. 
We plot  $N(E,{\mib r}_i)$ at the  $(1,1)$-site in the vortex core, 
and at the $(12,1)$-site far from the vortex.  
}
\label{fig:3}
\end{fullfigure}

The no-stripe case of Fig. \ref{fig:1} is shown in Fig. \ref{fig:3}(a). 
Far from the vortex,  $N(E,{\mib r}_i)$ shows the typical $d$-wave 
superconductor's DOS at zero field. 
The superconducting gap  $\Delta_0=2|V \Delta_{d,i}|\sim 0.96t$. 
At the vortex core, the superconducting gap is smeared, and 
a low-energy peak appears at $E\sim 0$. 
As for the stripe case of  Fig. \ref{fig:2}, Fig.  \ref{fig:3}(b) 
shows $N(E,{\mib r}_i)$ for the site within the stripe region, 
and  Fig.  \ref{fig:3}(c) for the site outside of the stripe. 
The metallic stripe region has more low-energy states than the outside region. 
Far from the vortex, the LDOS is reduced to that of the zero-field case.  
It is noted that there appears a small gap $\Delta_1$ ($\sim 0.2t$) 
within the $d$-wave superconducting gap 
$\Delta_0$ ($=2|V \Delta_{d,i}|\sim 0.6t$), 
because the $s$-wave component is induced in the $d$-wave 
superconductivity. 
This is also understandable from  the ARPES results,\cite{Zhou,Ino1,Ino2,Ino3}  
which report that the Fermi energy state near 
$(\frac{\pi}{2},\frac{\pi}{2})$ vanishes as a result of the stripe formation. 
Then, the low-energy state at the gap-node direction 
$(\frac{\pi}{2},\frac{\pi}{2})$ of $d$-wave superconductivity is absent 
in the superconducting stripe state,  
and a small gap opens in the $d$-wave superconducting gap. 
The structure of the peaks above $\Delta_0$ is due to the stripe structure. 
The vortex core state is completely different from that of the no-stripe case. 
There is no eminent low-energy state. 
At the vortex core, small peaks or shoulders appear above the small 
gap $\Delta_1$. 
They have similar structures to those of the STM 
results.\cite{pan,Renner,Maggio1,Maggio2}  

The vortex structure in the stripe state has characteristics 
in common with the Josephson vortex in layered superconductors 
under a parallel magnetic field. 
Since the superconductivity appears dominantly on the metallic 1D stripe 
lines, the interstripe coupling between the 1D superconducting stripes 
can be considered as Josephson-like coupling. 
Also, in the layered superconductors, the interlayer coupling is  
Josephson-like coupling, and the vortex is trapped in the interlayer 
space, where the superconductivity is weak. 
There, the coherence length around the vortex is long (short) parallel 
(perpendicular) to the layer. 
While there appears a zero energy state (precisely speaking, there is a 
small gap of the order $\Delta^2/E_{\rm F}$ in the $s$-wave pairing) 
in the vortex core state for a perpendicular field, 
the low-energy core state does not appear for a parallel field. 
The vortex core state has the energy of the order $\Delta_0$.\cite{IchiokaT} 
In contrast to the layered superconductors, 
the stripe line can move thermally. 
We sometimes need pinning centers in order to obtain the static stripe state, 
such as the partial substitution of Nd in LNSCO. 
The vortex core is a candidate for the pinning center to trap a stripe line.

In summary, we have studied the vortex structure in the superconducting 
stripe state, based on the Bogoliubov-de Gennes theory. 
The vortex state has a different structure from that of the 
usual no-stripe case. 
The vortex is trapped in the weak superconducting region outside of 
the stripe. 
The eminent low-energy quasiparticle states do not appear even at 
the vortex core, where only small peaks or shoulders appear above a 
small gap. 
These characteristics resemble those of the Josephson vortex 
in the layered superconductors under a parallel magnetic field.  
Although Figs. \ref{fig:2} and \ref{fig:3} represent the bond-centered 
stripe case, we observe the same type of behavior even in the 
site-centered stripe case.  

Our calculation assumes the static stripe order. 
Thus, precisely speaking, our results can be applied only to  
the static stripe state near $\frac{1}{8}$-filling. 
Apart from $\frac{1}{8}$-filling, we must consider the fluctuation 
effect of the stripe line, which will smear the structure of the static 
stripe case. 
However, we expect that the main characteristics such as the lack of 
an eminent low-energy electronic state will remain even in the fluctuating 
stripe case.  
Thus,  the effect of stripes should be taken into account 
when we analyze the exotic LDOS structure of the vortex state 
as observed by means of STM experiments.



\end{document}